# Van der Waals Heterostructures Based on Allotropes of Phosphorene and $MoSe_2$

Sumandeep Kaur[1,2*], Ashok Kumar[2*], Sunita Srivastava[1*] and K. Tankeshwar[3*]

[1]*Department of Physics, Panjab University, Chandigarh 160014, India*
[2]*Centre for Physical Sciences, School of Basic and Applied Sciences, Central University of Punjab, Bathinda, 151001, India*
[3]*Department of Physics, Guru Jambheshwar University of Science and Technology, Hisar, 125001, Haryana, India*

(June 19, 2017)

*e-mails:

Sumandeep Kaur (dusuman0015@gmail.com)

Ashok Kumar (ashok@cup.ac.in)

Sunita Srivastava (sunita@pu.ac.in)

K. Tankeshwar (tkumar@gjust.org)

# Abstract

The van der Waals heterostructures of allotropes of phosphorene (α- and β-P) with $MoSe_2$ (H-, T-, ZT- and SO-$MoSe_2$) are investigated in the framework of state-of-the-art density functional theory. The semiconducting heterostructures, β-P /H-$MoSe_2$ and α-P / H-$MoSe_2$, forms anti-type structures with type I and type II band alignments, respectively, whose bands are tunable with external electric field. α-P / ZT-$MoSe_2$ and α-P / SO-$MoSe_2$ form ohmic semiconductor-metal contacts while Schottky barrier in β-P / T-$MoSe_2$ can be reduced to zero by external electric field to form ohmic contact which is useful to realize high-performance devices. Simulated STM images of given heterostructures reveal that α-P can be used as a capping layer to differentiate between various allotropes of underlying $MoSe_2$. The dielectric response of considered heterostructures is highly anisotropic in terms of lateral and vertical polarization. The tunable electronic and dielectric response of van der Waals phosphorene/$MoSe_2$ heterostructure may find potentials applications in the fabrication of optoelectronic devices.

## 1. Introduction

Phosphorene with its semiconducting nature is newly included member in the family of two dimensional (2D) materials which is a potential candidate for nano- and opto-electronic applications due to its exceptional properties such as strong structural anisotropy, high mechanical strength, large lateral flexibility and high conductivity [1, 2]. The black phosphorene (α-P) exhibit puckered structure [3], whereas blue phosphorene (β-P) possess graphene-like honeycomb structure of phosphorous atoms [4]. Both α- and β-P are semiconductors with band gap ~1eV and ~2eV, respectively [4, 5]. Transition metal dichalcogenides (TMDs) are another family of 2D materials [6] with tunable electronic and dielectric properties [7-9]. The monolayers of TMDs are triatomic, in which the metal atoms are sandwiched between two atomic planes of chalcogen atoms. $MoS_2$ is a prototype material which is most explored among the TMDs [10].

Monolayer $MoSe_2$ is also emerging as a novel contender in electronic applications similar to $MoS_2$ with the fabrication of few-layer $MoSe_2$ based FETs having Ni and Ti electrode contacts possessing field-effect mobility of 150−200 $cm^2/(V·s)$ and a current on/off ratio up to $10^6$ [11, 12]. The layered crystal structure of $MoSe_2$ is found to exhibit two phases i.e., the trigonal prismatic phase (H-$MoSe_2$) and octahedral coordinated phase (T-$MoSe_2$), with 3 atoms in a hexagonal unitcell. The Se atoms lie directly above one another in H-$MoSe_2$, while they are displaced in case of T-$MoSe_2$. Another stable allotrope of $MoSe_2$ can be formed by repeating square octagon pairs in a square unitcell, which consists of 12 atoms per unitcell [13].

The absence of dangling bonds in 2D materials make it possible to reassemble the layers of these materials to form van der Waals heterostructures [14-17]. This layer by layer structure formed by the combination of two or more different materials, preserve the properties of these materials without much degradation and in addition exhibit some unusual phenomena too [18, 19]. The phosphorene based heterostructures have been used to develop novel devices for nanoelectronic applications in recent past e.g. STM images of h-BN/phosphorene heterostructure with h-BN as a capping layer is useful to distinguish between different structural phases of phosphorene [20]; phosphorene/$MoS_2$ acts as CMOS inverter [21] with its possible use as a channel material for electronic applications [22]; phosphorene/graphene heterostructure display potential application as electrode materials for batteries [23]; metal/black-phosphorene heterostructures such as Zn/BP and In/BP forms Schottkey contacts while Cu/BP forms ohmic contact [24] in devices. Similarly, heterostructures based on $MoSe_2$ are also found to be suitable for device applications such as the $MoSe_2$/graphene heterostructures show rapid transfer of photogenerated charge carriers between $MoSe_2$ and graphene which depicts its possible applications in optoelectronics [25]; black phosphorene/$MoSe_2$ heterostructure show potential applications in p-n diodes and logical devices [26];etc.

In the present work, we construct the novel commensurable heterostructures of black and blue phosphorene with various phases of monolayer $MoSe_2$ which show variable electronic structure on the influence of external electric field. Mechanical properties and simulated STM images of given heterostructures are also investigated. Obtained dielectric response is highly dependent on the direction of polarization.

## 2. Computational Method

SIESTA simulation package have been used to perform all the calculations [27]. The fully separable Kleinman and Bylander form of norm- conserving Troullier Martin pseudopotential has been used to treat the electron-ion interactions [28]. The van der Waals (vdW)-DRSLL functional have been employed in the present work to treat the exchange and correlation energies [29]. The Kohn Sham orbitals are expanded as a linear combination of numerical pseudo atomic orbitals using a split-valence double zeta basis set with polarization functions (DZP). Throughout geometry optimization, the confinement energy of numerical pseudo-atomic orbitals is taken as 0.01 Ry. Minimization of energy was carried out using standard conjugate-gradient (CG) technique. Structures were relaxed until the forces on each atom were less than 0.01 eV/Å. Monkhorst-Pack scheme is used to sample Brillouin zone with a 10×10×1 mesh for the calculations based on hetero-bilayers and a 30×30×1 mesh for homo-bilayers. The spacing of the real space used to calculate the Hartree exchange and correlation contribution of the total energy and Hamiltonian was 450 Ry. A vacuum region of about 20 Å perpendicular to 2D plane has been used in calculations to prevent the superficial interactions between the periodic images. The dielectric properties have been investigated by calculating the dipolar transition matrix elements between occupied and unoccupied single-electron eigenstates by means of first order time-dependent perturbation theory as implemented in SIESTA [27]. The static dielectric constant ( $\varepsilon_s$) is calculated as the value of the real part of the dielectric function at zero frequency, while electron energy loss spectra (EELS) is obtained from dielectric functions ($\varepsilon_1$and $\varepsilon_2$) using the formula:

$$im\left(-\frac{1}{\varepsilon(\omega)}\right)=\frac{\varepsilon_2(\omega)}{\varepsilon_1^2(\omega)+\varepsilon_2^2(\omega)}$$

## 3. Results and Discussions

The commensurable vertical heterostructures of two phosphorene allotropes namely α-P and β-P, are constructed with four phases of $MoSe_2$ namely hexagonal (H-$MoSe_2$), tetragonal (T-$MoSe_2$), distorted tetragonal (ZT-$MoSe_2$) and square octagon (SO-$MoSe_2$) (Figure 1). Note that ZT-phase of $MoSe_2$ is the distorted phase which is obtained after full structural optimization of the rectangular unitcell of T-$MoSe_2$ (Figure S1 of ESI), which is found to be stable in previous studies on $MoS_2$ [30].

Our calculations reveal that ZT-$MoSe_2$ is energetically favorable over T-$MoSe_2$ with an energy difference of 0.17 eV/atom**.**

The supercells of the vertical heterostructures are so chosen that the lattice mismatch between the layers get minimized e.g., α-P / H-$MoSe_2$ consists of 1 x 5 unitcells of α-P stacked vertically over 1 x 4 unitcells of H-$MoSe_2$ that leads to a minimum lattice mismatch of 3.2 %. The lattice mismatch in other heterostructures is calculated as 1.16 %, 6.7 %, 1.08% and 0.98 % for α-P / ZT-$MoSe_2$, α-P / SO-$MoSe_2$, β-P / H-$MoSe_2$ and β-P / T-$MoSe_2$, respectively. The equilibrium interlayer distance is obtained as 3.8 Å, 3.6 Å, 3.7 Å, 3.8 Å and 3.5 Å for α-P / H-$MoSe_2$, α-P / ZT-$MoSe_2$, α-P / SO-$MoSe_2$, β-P / H-$MoSe_2$ and β-P / T-$MoSe_2$, respectively. Binding energy per atom in these heterostructures is calculated to be a few meV (~ 40 meV) which shows weak van der Waals interaction in the considered heterostructures. Note that binding energy is obtained as $E_B = |E - (E_1 - E_2)|/N$, where E is the total energy of the heterostructure, $E_1$ and $E_2$ are the energies of the isolated individual layers of the heterostructures and N is the number of atoms in the heterostructure.

### 3.1 Electronic Structure

α-P displays a direct band gap of 1.12 eV while β-P has an indirect band gap, 2.05 eV (Table 1). H-$MoSe_2$ is a semiconductor with a direct band gap of about 1.13 eV while T-$MoSe_2$, ZT-$MoSe_2$ and SO-$MoSe_2$ are zero band gap materials. Note that GGA-PBE band gap underestimates the fundamental gap due to approximate nature of exchange-correlation functional. The band gap calculated with quasiparticle GW calculations in other reports [31-33] is ≈ 1.0 eV higher than that of GGA-PBE value for monolayer black phosphorene and $MoSe_2$. Therefore, it should be noted that the results presented here represent the qualitative picture of electronic band structure. ZT-$MoSe_2$ and SO-$MoSe_2$ have been found to exhibit Dirac-cone like feature at the Fermi level which arises due to the 4p orbitals of Se in ZT-$MoSe_2$ and 4d-orbitals of Mo in case of SO-$MoSe_2$ (Figure S2 of ESI).

We found that α-P/H-$MoSe_2$ and β-P/H-$MoSe_2$ are semiconductors with a band gap of 0.7 eV and 1.1 eV, respectively, former exhibit an indirect gap while the later has a direct band gap. In α-P / H-$MoSe_2$, the valance band maximum (VBM) is mainly contributed by 3p orbitals of P atoms whereas the conduction band minimum (CBM) is mainly composed of 4d-orbitals of Mo. In case of β-P / H-$MoSe_2$, the both VBM and CBM are mainly contributed by the 4d-orbitals of Mo (Figure 2). On the other hand, rest of the three heterostructures i.e., α-P / ZT-$MoSe_2$, α-P / SO-$MoSe_2$ and β-P / T-$MoSe_2$, are found to be zero-gap materials (Figure 2). On comparing the electronic structure of α-P / ZT-$MoSe_2$, α-P / SO-$MoSe_2$ with their monolayer counterparts, the Dirac-cone like feature are found to remains intact at the Fermi level while the P-bands shift upward. In β-P / T-$MoSe_2$, the bands at the Fermi level which arises due to T-$MoSe_2$, remains unaltered but the β-P bands show downward shift (Figure 2, Figure S2 of ESI). In all the heterostructures, the majority of the contribution to the total density of states around the Fermi level comes from the 4d, 4p and 3p orbitals of Mo, Se and P respectively (Figure S3 of ESI). In order to get further insight into the electronic structure, the charge density difference profile were obtained, which shows the accumulation of charge in van der Waals's gap that leads to reduction of energy gap due to interlayer interactions (Figure 2) as compared to higher band gap in the constituent semiconducting monolayers. Note that charge density difference is calculated as $\Delta\rho = \rho - (\rho_1 - \rho_2)$, where $\rho$ is the charge density of the heterostructure, $\rho_1$ and $\rho_2$ are the charge densities of the isolated monolayers.

**3.2 Band Alignment:**

We found that *p*-type phosphorene and *n*-type H-$MoSe_2$ forms an anti-type heterostructure with type I and type II band alignments for β-P / H-$MoSe_2$ and α-P / H-$MoSe_2$, respectively (Figure 3). The type II heterostructures are ideal for solar energy conversion and optoelectronic applications since the free electrons and holes get instinctively separated [22]. There is a small amount of band bending as one moves from the region where the phosphorene and $MoSe_2$ layers are stacked to the region where only pristine phosphorene is present. Note that band bending is obtained as the difference between the work functions of the composed system (W) and the pristine layer ($W_s$): $\Delta E_F = W - W_s$. The magnitude of band bending in α-P / H-$MoSe_2$ is found to be 0.17 eV for α-P and 0.11 eV for H-$MoSe_2$ while its magnitude in β-P / H-$MoSe_2$ comes out to be 0.07 eV and 0.43 eV for β-P and H-$MoSe_2$, respectively. Similarly, the values of band bending for phosphorene layer in α-P / ZT-$MoSe_2$, α-P / SO-$MoSe_2$ and β-P / T-$MoSe_2$ are 0.53 eV, 0.60 eV and - 0.62 eV, respectively (Figure 3). Since $\Delta E_F > 0$ for α-P / ZT-$MoSe_2$ and α-P / SO-$MoSe_2$, electrons will be transferred from phosphorene to $MoSe_2$ making it as a *p*-type channel while for β-P / T-$MoSe_2$ the semiconductor-metal contact is *n*-type due to negative band bending. Note that when two layers with different work functions comes in contact with each other, the charges redistribution and bend bending occurs to equalize the work functions between the layers. Also in case of vertical heterostructures, the weak interlayer coupling results in the change in band positions of the constituent monolayers in heterostructure.

It is important here to discuss the Schottky Barrier Height (SBH) for the semiconductor- metal heterostructures if we consider $MoSe_2$ as a metal contact and phosphorene as the channel material for device application. The Schottky barrier in a semiconductor- metal junction is the excitation energy of electrons from metal to semiconductor or vice-versa. For *p*-type semiconductors, the Schottky barrier is defined as the difference between the Fermi level and the valance band edge of the semiconductor and for *n*-type materials it is the difference between the conduction band edge of the semiconductor and the Fermi level. In α-P / ZT-$MoSe_2$, α-P / SO-$MoSe_2$ and β-P / T-$MoSe_2$, the SBH is calculated as 0 eV, 0 eV and 0.5 eV, respectively. Hence, two of the heterostructures namely α-P / ZT-$MoSe_2$ and α-P / SO-$MoSe_2$ results in an ohmic contact. Hence, all three of our semiconductor-metal contacts owning very low Schottky barrier height, if employed for practical applications, can be proved to greatly enhance the drive current of the device. The Schottky Barrier is *p*-type for α-P / ZT-$MoSe_2$ and α-P / SO-$MoSe_2$ since the valance band is closer to the Dirac point while it is *n*-type for β-P / T-$MoSe_2$ (Figure 2).

### 3.3 Effect of External Electric Field:

On applying perpendicular electric field in the range ± 1 **V/Å**, the semiconducting heterostructures show reduction in band gap, however, change in band gap depends strongly on the polarity of the field e.g positive field reduces the bandgap sharply leading to a transition from semiconductor to metal at $E_\perp$ = 0.5 V/Å in α-P / H-$MoSe_2$ while negative field firstly increases the band gap to a maximum value at $E_\perp$ = 0.2 **V/Å** and then decreases continuously to minimum (Figure 4). Similarly, β-P / H-$MoSe_2$ show asymmetric behavior in band gap variation with respect to the polarity of applied external electric field that results into direct-to-indirect bandgap transition. The change in bandgap on the application of external electric field is as a result of charge redistribution in van der Waal's gap as shown for α-P/H-$MoSe_2$ in Figure 5. The asymmetric behavior of the bandgap with electric field in case of semiconducting heterostructures can be attributed to the counterbalance of external electric field with the internal electric field arising due to the difference in electronegativity and structural arrangements of the atoms in the two layers of the heterostructure. The evolution of bands as a function of applied electric field is given in Figure 5 and Figure S4 of ESI.

In case of metallic heterostructures, the tuning of Schottky barrier height (SBH) with applied electric field can be achieved. Particularly, on the application of positive electric field in α-P / ZT-$MoSe_2$ and α-P / SO-$MoSe_2$, the semiconductor-metal contact remains ohmic but the Dirac cone moves energetically below the valence band edge of α-P making the SHB negative while for negative fields, the Dirac cone remains at the Fermi level but the Schottky barrier changes from *p*-type to *n*-type (Figure 5 and Figure S4 of ESI). In case of β-P / T-$MoSe_2$, however, positive field turns the barrier from *n*-type to *p*-type as the valence band edge of β-P becomes closer to the Fermi level while negative fields of around – 1.0 **V/Å** reduces the barrier height to zero rendering the given heterostructure to possess ohmic contact. Obtaining zero or even very low SBH in device contacts is one of the most-crucial tasks for realizing high-performance nano-material-based FETs [34, 35].

The change in electronic structure with applied field can be quantified in terms of charge density difference profiles. On the application of positive electric field, an accumulation of electrons occur near the $MoSe_2$ layer while the region near phosphorene layer appears to be depleted of charge which clearly depicts electron transfer from phosphorene to $MoSe_2$ layer. Similarly, on applying negative field, electrons (holes) tend to move from $MoSe_2$ (phosphorene) to phosphorene ($MoSe_2$) (Figure 5).

**3.4 Mechanical Properties:**

The elastic parameters such as in-plane stiffness and Poisson's ratio are used to characterize the mechanical response of a two-dimensional material. In-plane stiffness is given by $C = \frac{1}{S_0}\left(\frac{\partial^2 E_S}{\partial \epsilon^2}\right)$ where $S_0$ is the equilibrium area of the two-dimensional material, $E_s$ is the strain energy and ϵ is the applied strain, whereas the Poisson's ratio is defined as $\nu = \frac{-\epsilon_{trans}}{\epsilon_{axial}}$ [36, 37]. A strain ranging from -0.02 to 0.02 is applied by changing the lattice constant along x and y directions, simultaneously, in steps of 0.01 to make a data grid of 25 points to calculate the strain energies which were then fitted in the formula $E_s = a\epsilon_x^2 + b\epsilon_y^2 + c\epsilon_x\epsilon_y$ where $\epsilon_x$ and $\epsilon_y$ are strain along $x$ and $y$ directions, respectively. Thereafter, the in-plane stiffness and Poisson's ratio along $x$ direction is obtained as $C_x = \frac{1}{S_0}\left(2a - \frac{c^2}{2b}\right)$ $and\ v_x = \frac{c}{2b}$ respectively while along $y$ direction these are obtained to be $C_y = \frac{1}{S_0}\left(2b - \frac{c^2}{2a}\right)$ $and\ v_y = \frac{c}{2a}$.

In case of hexagonally symmetric systems, $a$ and $b$ are equal in harmonic range. Therefore, the in-plane stiffness and poisson's ratio of H-$MoSe_2$, T-$MoSe_2$ and blue-P are same along x and y directions. Also for the squared symmetric SO-$MoSe_2$, it is same in both the directions. On the other hand, the anisotropy in ZT-$MoSe_2$ and black-P structures along x and y directions lead to different in-plane stiffness and poisson's ratio along x and y directions (Table 1). A strong bond (i.e., with smaller bond length) and higher bond density leads to higher in-plane stiffness. But the average bond lengths along x and y directions in all the considered $MoSe_2$ phases is almost the same and also the Mo atoms in all the structures are six fold coordinated. Hence the difference in in-plane stiffness of these phases may be attributed to the different structural arrangements of Mo-Se units in different phases. SO-$MoSe_2$ and β-P are found to be the most and least flexible layers with smallest (~ 46 N/m) and highest (~ 277 N/m) values of in plane stiffness, respectively. Also, the highest values of Poisson's ratio for SO-$MoSe_2$ (0.63) suggests its ease in expansion without braking.

The in-plane stiffness and Poisson's ratio of the considered heterostructures is also depicted in Table 1. In-plane stiffness values of considered heterostructures follow the trend β-P / H-$MoSe_2$ > β-P / T-$MoSe_2$ > α-P / ZT-$MoSe_2$ > α-P / H-$MoSe_2$ > α-P / SO-$MoSe_2$. Hence, α-P / SO-$MoSe_2$ with lowest in-plane stiffness and highest poisson's ratio among all the heterostructures, may find application in flexible van der Waals device fabrication.

**3.5 Simulated STM Images:**

First principles calculations have been well known to simulate the STM images by applying a small bias voltage $V_{bias}$ between the STM tip and the sample [20, 38-40] which yields a tunneling current with density $j(r)$ proportional to the local density of states (LDOS) at the centre of the curvature of the tip at (r), approximated by the Tersoff and Hamann [41]. The LDOS is integrated from ($E_F - eV$) to $E_F$ and the *s*-waves with constant density of states describes the tip states. A positive (negative) bias voltage indicates tunneling from occupied states of the sample to tip (tip to unoccupied states of the sample). WSxM code [42] has been employed hereto obtain the STM images derived from LDOS in constant current (CC) mode. STM images of α-P containing heterostructures are simulated at ± 0.5 V while for β-P based heterostructures images are obtained at higher bias i.e. ± 1.5 V due to higher energy gap in the constituent monolayer of phosphorene, so as to get a better understanding of the electronic states near the Fermi level. Simulated STM images of the constituent monolayers are clearly distinguishable due different atomic arrangement in the given structures (Figure 6). Bright colors in all the STM images indicate the atoms nearest to the STM tip i.e., the top most atoms. In case of α-P, for forward bias, the electron density is large over the P atoms while for reverse bias the electron density is localized over the covalent bonds between the pair of P atoms. This means an antibonding character is visible for positive bias while bonding character is visible for negative bias. In case of β-P, under forward as well as reverse bias the contribution to electronic density is only from upper P atoms (Figure 6 and Figure S5 of ESI). Similar explanation is applicable for understanding the electronic density distribution of other structures as well.

On comparing STM images of heterostructures with that of the corresponding monolayers, we observe that in addition to possessing the antibonding character of phosphorous in α-P based heterostructures, the features of $MoSe_2$ are clearly depicted. Therefore, α-P can be potentially use as a capping layer to differentiate between different allotropes of $MoSe_2$. On the other hand, no difference in STM images was obtained for β-P based heterostructures (Figure 6 and Figure S5 of ESI), hence β-P cannot be used as a capping layer to differentiate between different $MoSe_2$ allotropes.

### 3.6 Dielectric Properties:

Now we calculate the dielectric response of given heterostructures and their constituent monolayers (Figure 7 and Figure S6 of ESI). The dielectric functions are found to be highly anisotropic in low energy range and become isotropic in high energy range. The structure peaks in the imaginary part of dielectric function ($\varepsilon_2$) is associated with the interband transitions in the corresponding electronic structure. Note that, there are multiple interband transitions due to different band numbers and different brillouin zone directions that might corresponds to a particular structure peak in the imaginary part of the dielectric function. For example, α-P and β-P possess two prominent structure peaks one at ~ 4 eV corresponding for lateral polarization and the other at ~ 6 eV due to the vertical polarization which corresponds to interband transition between the bands separated by ~ 4 eV and ~ 6 eV across the Fermi level in corresponding electronic structure which falls in the ultraviolet region (3.2 eV to 12 eV). The energy of the structure peaks remain nearly same for the homobilayers as well but the intensity increases (Figure S6 of ESI). The higher intensity in case of homobilayers is attributed to the availability of increased number of allowed bands for interband transition. In case of heterostructures also, the interband transition is found to occur in ultraviolet region and the prominent feature is mainly due to phosphorene (Figure 7).

The structure peaks in electron energy loss spectra (EELS) depicts the plasmonic excitations. Low energy plasmonic structures (π-plasmons) are related to the excitation of weak π-electrons while high energy plasmonic features ( π + σ plasmons) are associated with both weak π- and strong σ-electrons [43]. For example, α-P exhibits two plasmonic features, one at lower energy < 8 eV and other at energy > 8 eV (Figure 7). Note that five valence electrons in phosphorous atom gives rise to $sp^3$ hybridization, three of the electrons form covalent bonds with other three atoms while rest of the two electrons occupy lone pair. Similarly, other monolayers considered in the study also exhibits two types of plasmonic features. EELS of homobilayers case also exhibit two plasmonic peaks which get blue shifted with respect to their monolayer counterparts (Figure S6 of ESI).

In case of heterostructures, the EELS of α-P / H-$MoSe_2$, α-P / ZT-$MoSe_2$ and α-P / SO-$MoSe_2$ depicts high energy plasmonic feature corresponding to π- and σ-electrons while the EELS of β-P / H-$MoSe_2$ and β-P / T-$MoSe_2$ exhibits both high and low energy plasmonic feature (Figure 7). The calculated values of static dielectric constant ($\varepsilon_s$) shows highly anisotropic response of given monolayers and heterostructures e.g. the value of $\varepsilon_s$ for α-P is 3.7 and 2.2 for lateral and vertical polarization, respectively, while these are 7.6 and 4.2 for α-P / H-$MoSe_2$ heterostructure. The value of $\varepsilon_s$ in semiconducting monolayers and heterostructures show inverse trend in comparison with the corresponding band gap which can be attributed to the two band relationship $\varepsilon_s \approx 1 + \left(\frac{\hbar\omega_P}{E_g}\right)^2$ between the bandgap $E_g$ of the semiconductors and $\varepsilon_s$, proposed by Penn [44], where $\omega_P$ is the plasmonic frequency. Higher value of static dielectric constant represent the characteristics of metal [Table 1]. . The variable dielectric response of given heterostructures make them useful for tunable optoelectronic applications.

## 4. Summary:

In summary, first principles calculations have been employed to investigate various properties of phosphorene/$MoSe_2$ heterostructures. The commensurate phosphorene based heterostructures are found to be semiconducting and metallic in nature. α-P / ZT-$MoSe_2$ and α-P / SO-$MoSe_2$ are found to form ohmic contacts whereas β-P / H-$MoSe_2$ exhibit metallic features. External electric field induces semiconductor-to-metal transition in α-P / H-$MoSe_2$ while direct-to-indirect bandgap transition occurs in β-P / H-$MoSe_2$. In case of α-P / ZT-$MoSe_2$ and α-P / SO-$MoSe_2$, the positive field makes the schottky barrier height negative while negative field changes the barrier from *p*-type to *n*-type. In β-P / T-$MoSe_2$, positive field change the barrier from *n*-type to *p*-type while negative field changes the contact to ohmic. Mechanical response of given heterostructures reveal that α-P / SO-$MoSe_2$ can be potentially use as a material for flexible nanoelectronics. Simulated STM images of given heterostructures depict that α-P can be used as a capping layer to distinguish various allotropes of $MoSe_2$ . The dielectric response of considered heterostructures is found to be highly anisotropic. Our results demonstrate that considered structures have great potential to be used as novel materials for nano-electronic device applications based on van der Waal heterostructures.

## Acknowledgements:

SK is grateful to UGC-BSR for financial assistance in terms of junior research fellowship. We acknowledge computational facility provided by Physics department, Panjab University, Chandigarh. A part of work presented in the paper is performed at Central University of Punjab, Bathinda.

**TABLE 1.** Lattice parameter (a, b), band gap ($E_g$), in-plane stiffness, Poisson's ratio, lateral ($\varepsilon_S^{\parallel}$) and vertical ($\varepsilon_S^{\perp}$) components of static dielectric constant ($\varepsilon_s$) of the given monolayers and heterostructures.

| | **Monolayers** | | | | | | **Heterostructures** | | | | |
|---|---|---|---|---|---|---|---|---|---|---|---|
| **Parameter** | **H-$MoSe_2$** | **T-$MoSe_2$** | **ZT-$MoSe_2$** | **SO-$MoSe_2$** | **α-P** | **β-P** | **α-P / H-$MoSe_2$** | **α-P/ ZT-$MoSe_2$** | **α-P / SO-$MoSe_2$** | **β-P/ H-$MoSe_2$** | **β-P/ T-$MoSe_2$** |
| **a, b (Å)** | 3.4, 5.9 | 3.4, 6.1 | 3.38, 6.06 | 6.80, 6.80 | 3.41, 4.84 | 3.41, 3.41 | 3.41, 24.18 | 3.41, 24.18 | 6.83, 14.51 | 3.41, 3.41 | 3.41, 3.41 |
| **Supercell** | 1 x 1 | 1 x 1 | 1 x 1 | 1 x 1 | 1 x 1 | 1 x 1 | 1×5 / 1×4 | 1×5 / 1×4 | 2×3 / 1×2 | 1×1 / 1×1 | 1×1 / 1×1 |
| **$E_g$ (eV)** | 1.13 (2.26)[a] | 0 | 0 | 0 | 1.12 (1.94)[b] | 2.05 | 0.67 | 0 | 0 | 1.10 | 0 |
| **In-plane stiffness ($C_x$, $C_y$) (N/m)** | 107.0, 107.2 | 123.8, 123.7 | 111.90, 125.21 | 46.10, 46.10 | 81.60, 60.42 | 277.12, 276.48 | 181.44, 149.96 | 190.43, 184.23 | 125.74, 110.78 | 701.12, 703.04 | 675.36, 675.20 |
| **Poisson's ratio ($\nu_x$, $\nu_y$)** | 0.34, 0.34 | 0.09, 0.09 | 0.27, 0.30 | 0.63, 0.63 | 0.35, 0.26 | 0.02, 0.02 | 0.37, 0.30 | 0.32, 0.31 | 0.43, 0.38 | 0.14, 0.14 | 0.07, 0.07 |
| $\boldsymbol{\varepsilon_S^{\parallel}}$ | 4.6 | 44.6 | 24.7 | 17.4 | 3.7 | 3.0 | 7.6 | 23.9 | 19.3 | 6.5 | 47.4 |
| $\boldsymbol{\varepsilon_S^{\perp}}$ | 2.7 | 2.8 | 2.8 | 2.5 | 2.2 | 1.8 | 4.2 | 4.0 | 4.1 | 3.7 | 4.0 |

GW-bandgap from [a]Ref. [31], [b]Ref. [32]

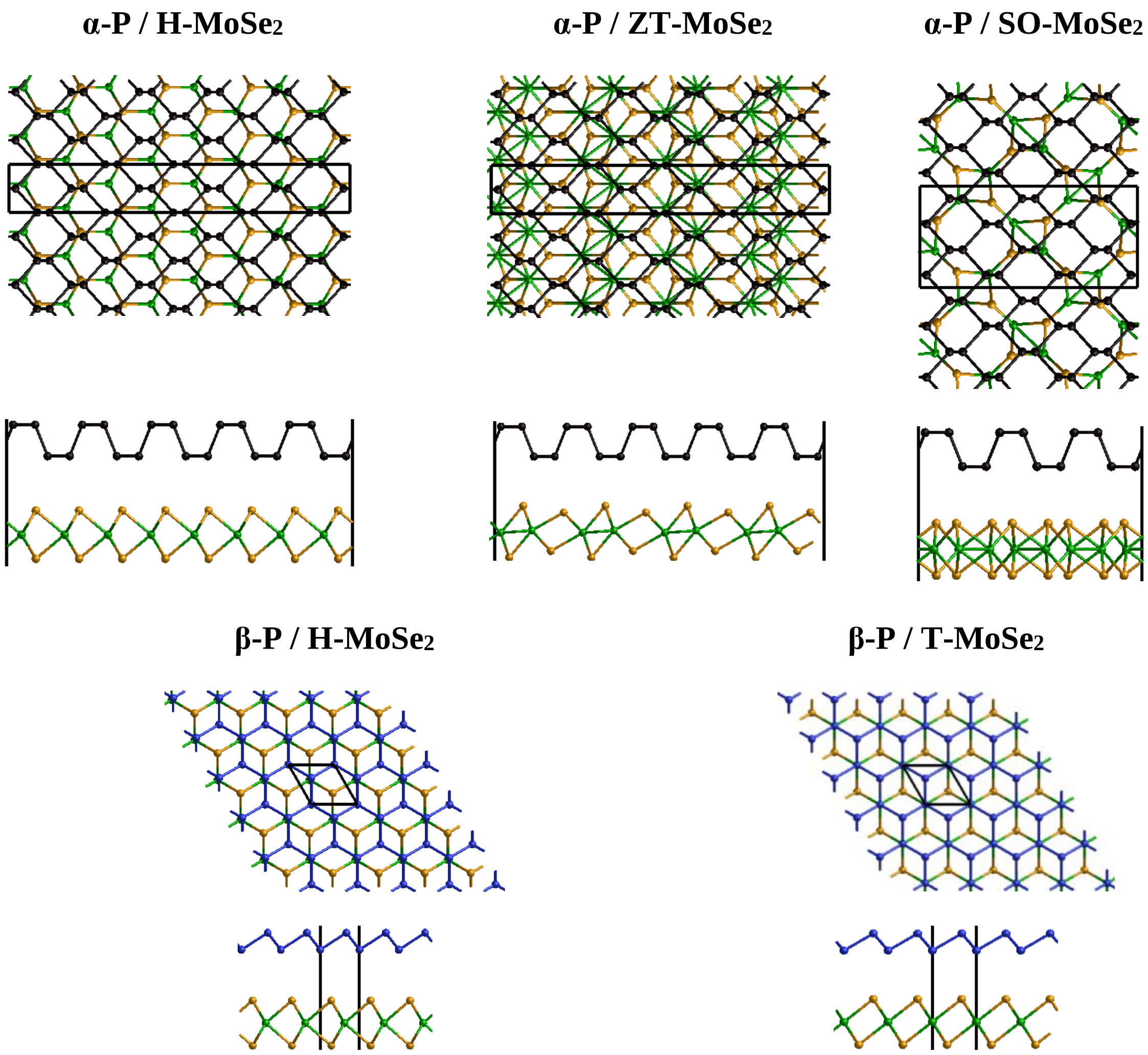

**FIGURE 1:** Top and side view of the commensurable heterostructures. Black and blue balls indicate P atoms while green and yellow balls indicate Mo and Se atoms respectively.

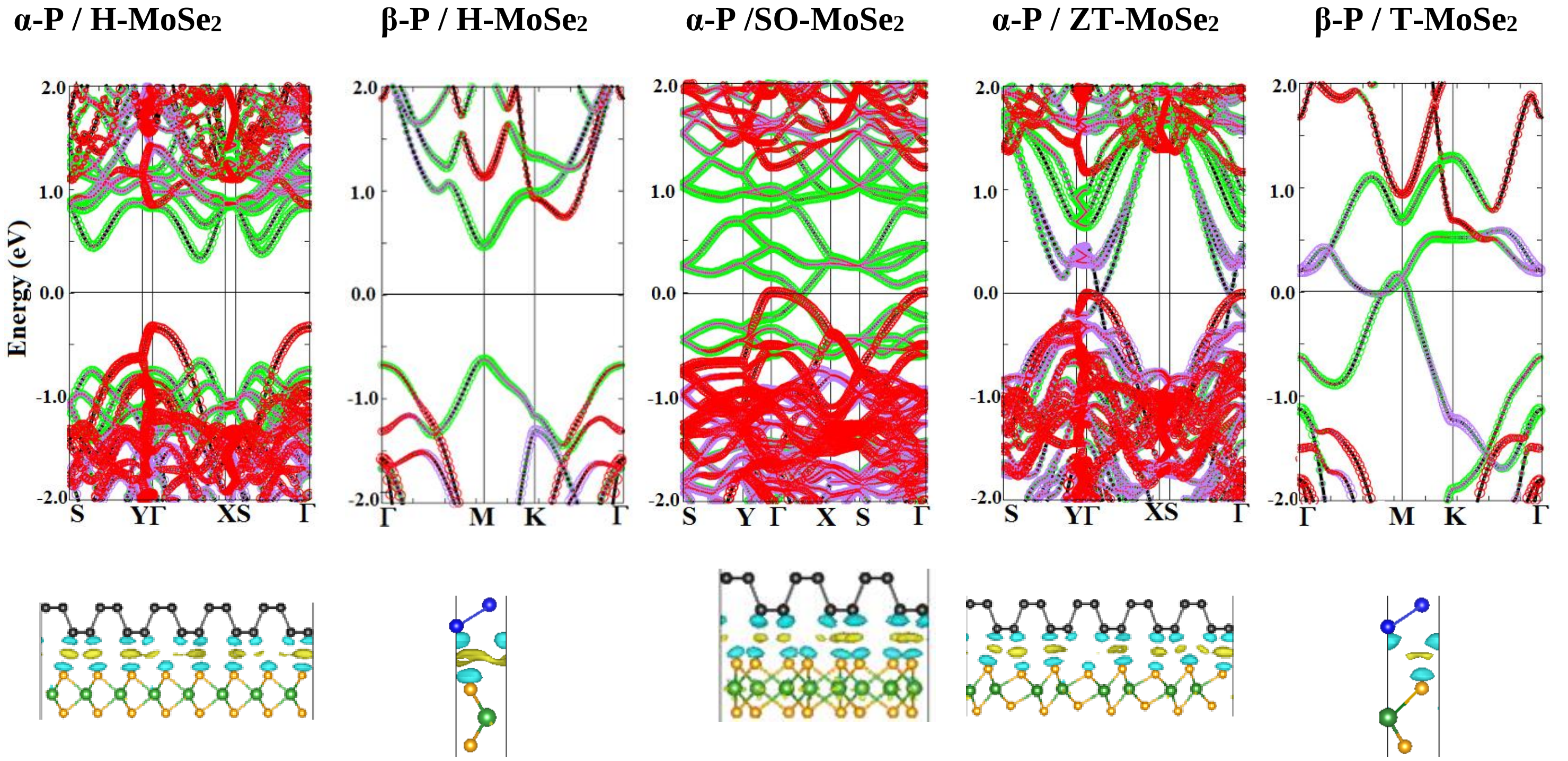

**FIGURE 2:** Orbital-resolved electronic band structure of heterostructures. Red, green and purple color in the band structure indicates the contribution of 3p orbitals of P, 4d orbital of Mo and 4p orbital of Se, respectively. The yellow and cyan color in charge density difference plots (lower panel) represent the accumulation and depletion of charges, respectively. Isosurface value for charge density difference plots is taken as $10^{-4}$ e/Å$^3$.

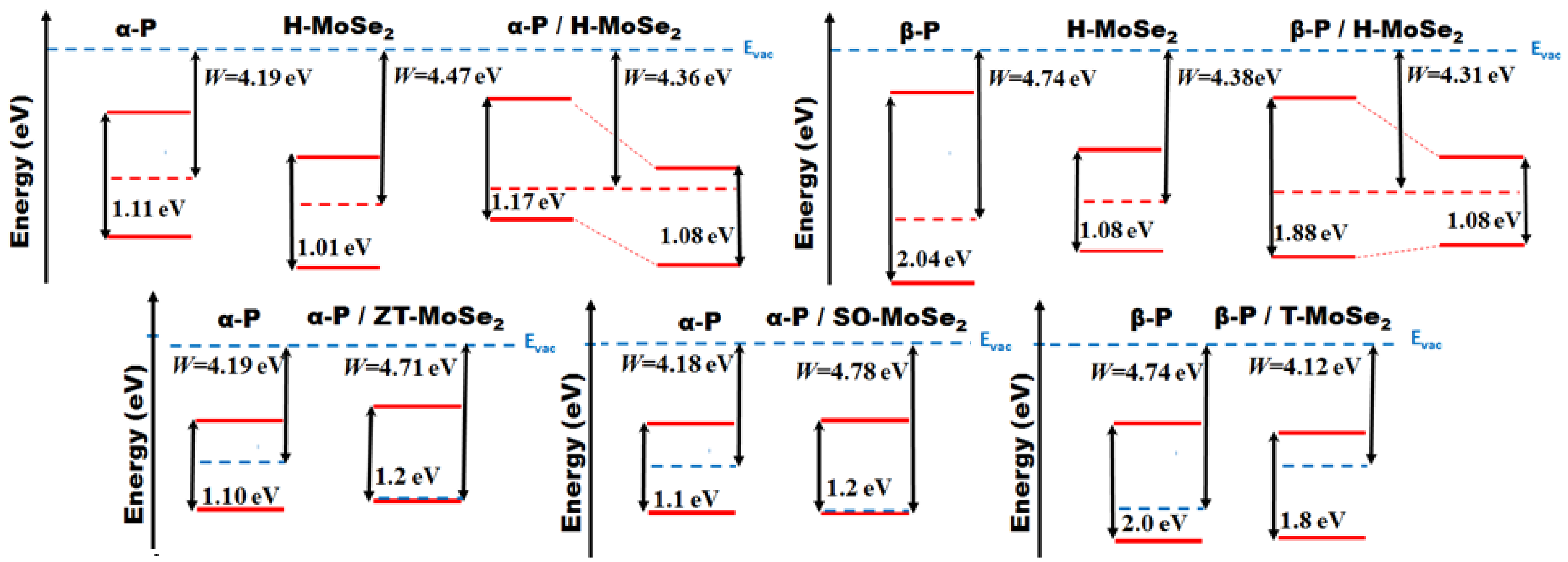

**FIGURE 3:** Band alignments of the given heterostructures. Dashed line between solid red lines is Fermi level. W and $E_{vac}$ indicates the work function and vacuum energy levels, respectively.

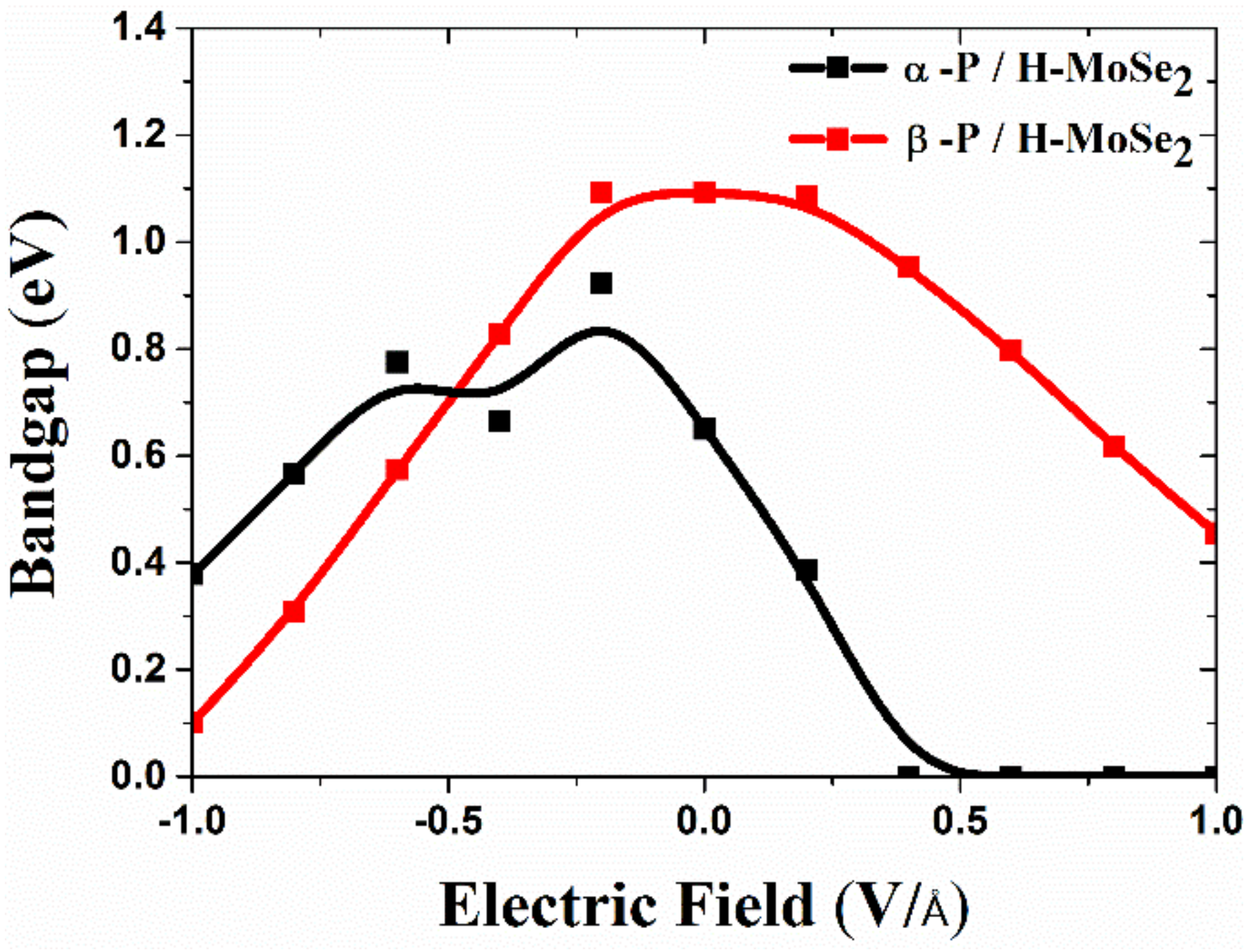

**FIGURE 4:** Band gap variation with applied electric field for semiconducting heterostructures.

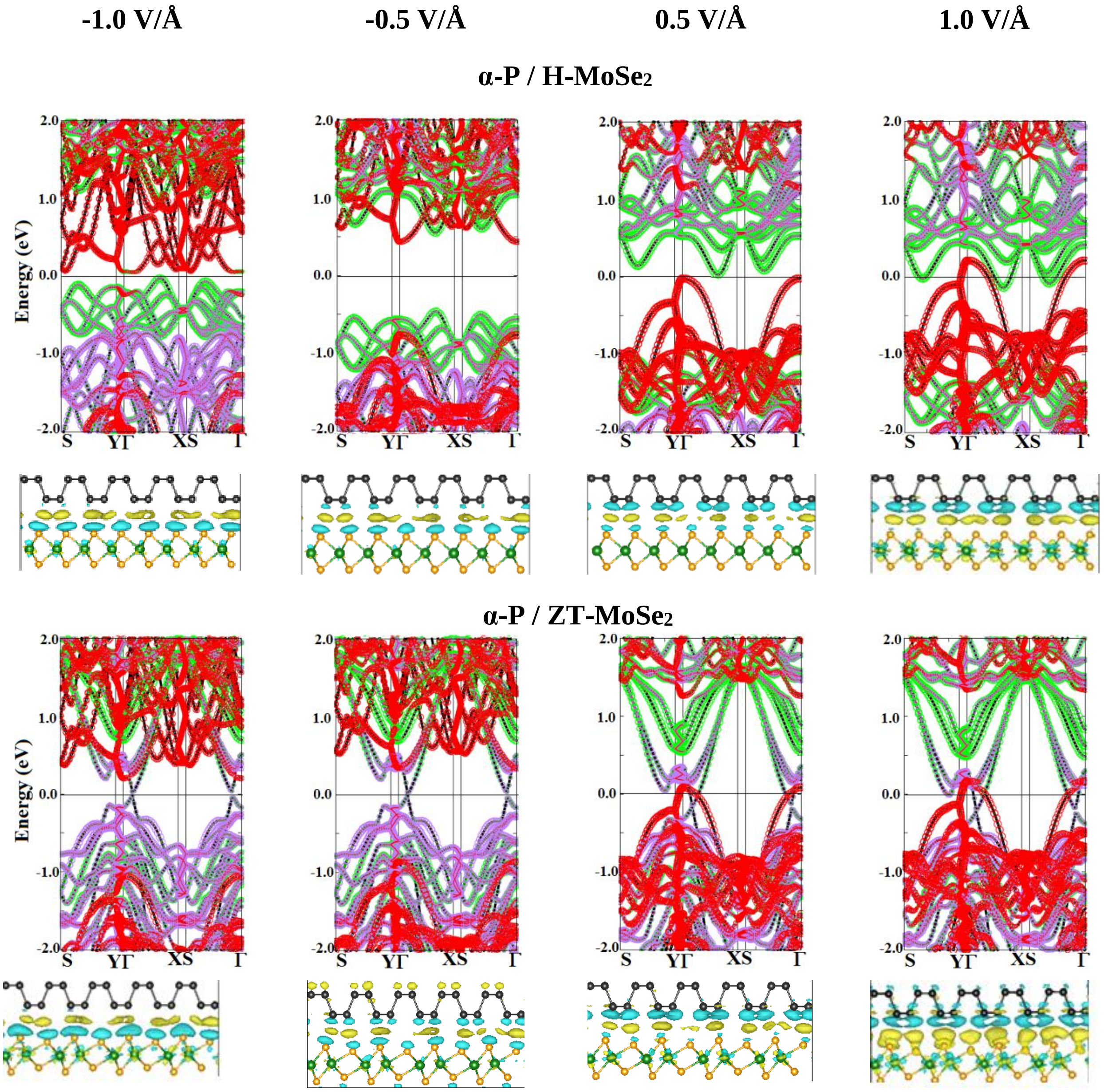

**FIGURE 5:** Bandstructures and charge density difference plots for semiconducting (α-P / H-$MoSe_2$) and metallic (α-P / ZT-$MoSe_2$ ) heterostructure at different applied electric field. Red, green and purple color in the band structure indicates the contribution of 3p orbitals of P, 4d orbital of Mo and 4p orbital of Se respectively. Isovalue for charge density difference plots is taken as $10^{-4}$ e/Å$^3$.

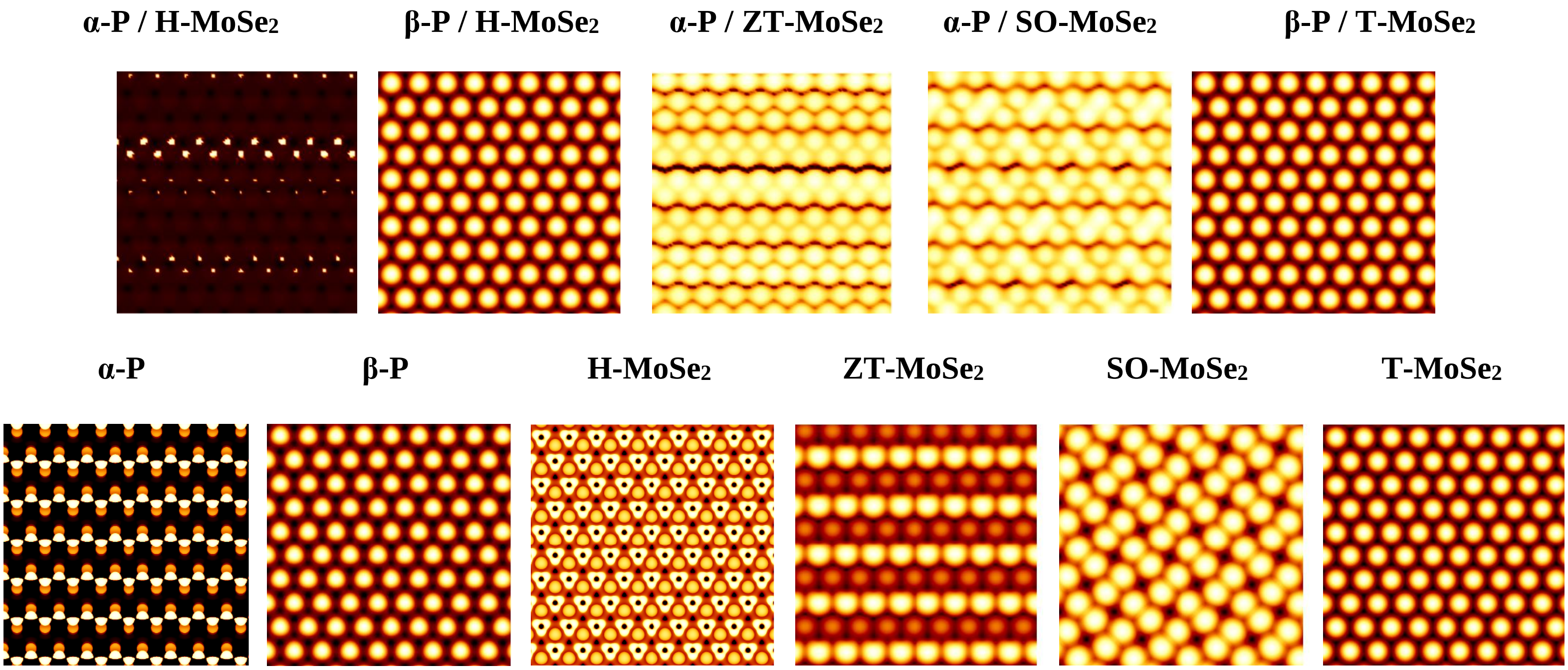

**FIGURE 6.** Simulated STM images of dimension 30 Å × 30 Å for various heterostructures compared with the monolayers. The isosurface is taken at 1.0 $\times 10^{-6}$ e/Å$^3$.The bias voltage for α-P based heterostructures is + 0.5 V and for β-P based heterostructures it is +1.5 V.

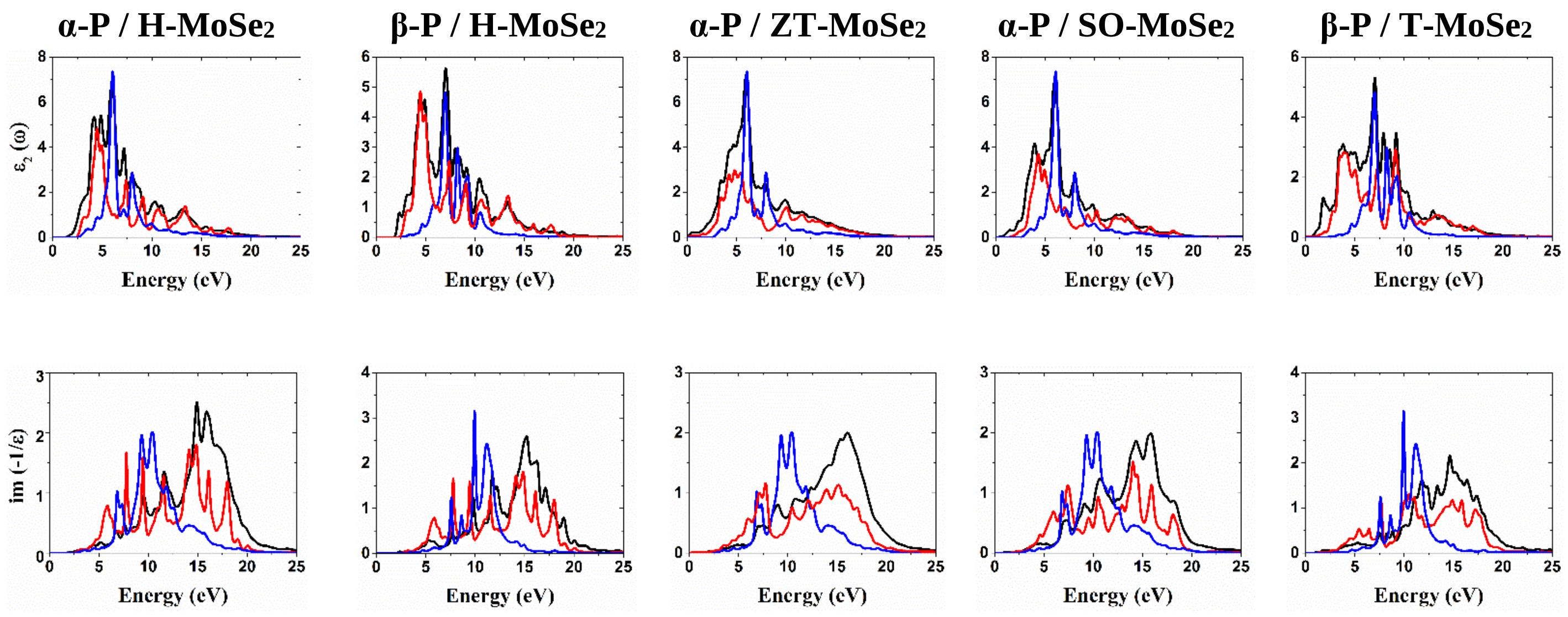

**FIGURE 7.** Imaginary part of the dielectric function ($\varepsilon_2$) and electron energy loss spectra [im (-1/ε)] of given heterostructures (black curve) for lateral polarization. Spectra of constituent's monolayers (red curve for $MoSe_2$ and blue curve for phosphorene) are also given for comparison.